# Surface phonons in two-layer thin films of GeSe

F.M. Hashimzade, D.A. Huseinova, Z.A.Jahangirli, B.H. Mehdiyev

Institute of Physics, National Academy of Sciences of Azerbaijan, AZ 1143, Baku, Azerbaijan

**Abstract.** This paper presents ab inition calculations of the surface phonon spectra of GeSe layered semiconductor compound, based on the Density Functional Perturbation Theory (DFPT). The surface has been imitated by a structure of periodically arranged slabs of two layers of GeSe crystal separated from other identical layers by the layers of vacuum sufficiently wide to ignore the effect of the upper boundary of the double-layer upon its lower boundary. We discuss the character of the surface modes located in the gaps, in the pockets, and in the area of allowed phonon states for the bulk GeSe crystals, as well as outside (above and below) the boundaries of the bulk phonon states.

**1.Introduction**
*GeSe* layered compound (along with *GeS*, *SnSe* and *SnS*) with strongly pronounced anisotropy of physical properties has interesting electronic, optical and dynamic properties. These compounds are promising materials for manufacturing of photodetectors and lasers in the near-infrared range, and are used as the cutoff devices and as the absorber materials in the non-toxic photovoltaic devices. Despite the fact that the electronic and optical properties of these compounds are well studied, the dynamic properties, particularly the dynamic surface properties, have remained virtually unexplored.

It is well known that various lattice defects and surface effects change electronic spectrum of a crystal. The presence of impurities and surface effects also drastically change the vibrational spectrum of a crystal. The existence of a surface can lead to the appearance of local and resonant states that are not present in a perfect crystal. Local and resonant states always affect the structure and the optical phonon spectra in semiconductors. Since real crystals contain the surface the effect of which cannot always be neglected, proper consideration of the effect of defects is essential in the analysis and interpretation of experimental data,. Furthermore, surface and interface structure are currently important working elements in many semiconductor devices.

The linear response formalism is known to yield reliable results for the lattice dynamics of both bulk and surfaces of a large variety of covalent crystals [1]. It has been shown that the density functional perturbation theory (DFPT) can be successfully applied to layered systems, such as *GeSe*, where the interlayer bonding is governed by van der Waals-type forces [2]. In this paper we present an ab initio calculation of the germanium selenide surface phonon structure based on DFPT.

**2. The crystal structure and calculation method**
*GeSe* has an orthorhombic structure and belongs to the $P_{nma}$ ($D_{2h}^{16}$) space group. Its lattice parameters are a=4.414 Å, b=3.862 Å and c=10.862 Å. The positions of the atoms in the structure, in fractional coordinates, are as the following : both atoms are at 4c as $\pm(x;1/4;z)$ and $\pm(1/2-x;3/4;1/2+z)$ [3]. A unit cell contains eight atoms organized in two adjacent double layers that are perpendicular to the direction of the c-axis. The atoms in each double layer bond to their three nearest neighbours by covalent bonds and form a zigzag chain along the direction of the minor axis of the crystal. As a consequence of the dominant van der Waals character of the bonds between adjacent layers, this material cleaves easily along the [001] planes.
All commonly used approaches to calculation of the frequencies of surface phonons are based on the slab method. Here we use the geometry of the periodically arranged slabs, wherein each slab comprises two layers. Adjacent slabs are separated by vacuum with the thickness of 13.72 Å. The thickness of each layer, which in turn consists of four atomic planes arranged in sequence Ge-Se-Se-Ge, equals 2.585 Å and the interlayer distance is 2.691 Å. Thus, the overall thickness of the GeSe slab is 7.86 Å. The model that we constructure for this superstructure also contains 8 atoms in a unit cell, but its symmetry is significantly lower, namely, the P21/m (No.11) space group. The calculations were performed in the DFPT framework based on the plane-wave pseudopotential method and implemented in the ABINIT program package [4].
The exchange-correlation interaction has been described in the local density approximation scheme [5]. For the norm-conserving pseudopotentials we used the Hartvigsen-Goedekker-Hutter (HGH) pseudopotentials [6]. In the expansion of the wave functions we included the plane waves with energies up to 40 Ha, which ensures a good convergence of the total energy. The equilibrium structure has been determined by minimizing the total energy relative to the lattice parameters, and the internal structure parameters have been optimized using Hellmann- Feynman forces . The minimization process was carried out until the force modules become less than 10-7 Ha/Bohr. The integration over the Brillouin Zone (BZ) has been carried out by $4\times 4\times 1$ partition shifted from the origin by (0.5,0.5,0.5), according to the Monkhorsta-Pak scheme [7]. We use the ANADDB routine from the ABINIT package for the Fourier transforms to construct the dispersion law of the phonon modes over the entire BZ; this allows calculating phonon frequencies in arbitrary points of the BZ. In the calculated density of the phonon states one can see the resonances, the anti-resonances, and the gap and localized states.

**3. Discussion of the results**
In addition to the localized surface states and the resonant surface zones, the phonon-dispersion curves of a crystal with surface contains the bulk zones. The latter correspond to the eigenstates that are delocalized in the direction perpendicular to the surface, and their energies correspond to the energies of states in an infinite bulk crystal. To determine the phonon spectrum of a crystal with surface one has to project the zones of a bulk crystal onto the first surface BZ.
The calculated projected bulk dispersion curves, the surface modes and the mixed modes along symmetrical directions of the BZ for GeSe are shown in Fig. 1. In the projected zone structure one can see energy gaps and energy pockets where the localized states are located.
An important feature of the spectrum of the surface phonon states is the appearance of high-frequency surface modes above the optical continuum at point **Γ**. Since the surface atoms have fewer nearest neighbours for bonding, the restoring forces and, consequently, the force constants on the surface are smaller than in the bulk. Therefore, one should expect that the surface modes will have lower frequencies compared to the bulk modes. In our case the surface phonons have higher frequencies than the bulk ones. This can be explained by the fact that in GeSe on (001)-surface during the relaxation of the surface atoms the bonds between the surface atoms change, and, as a result, the effective force constant increase. Indeed, the results of the calculations of atomic relaxations show that near the border with vacuum the lengths of the "vertical" *Ge-Se* bonds (2.48 Å) are sigificanlty shorter than the bond length in the bulk (2.51 Å) whereas the "horizontal" bonds almost do not change (2.58 Å). Therefore, the surface phonon modes appear above the optical bulk continuum as shown in Fig. 1.

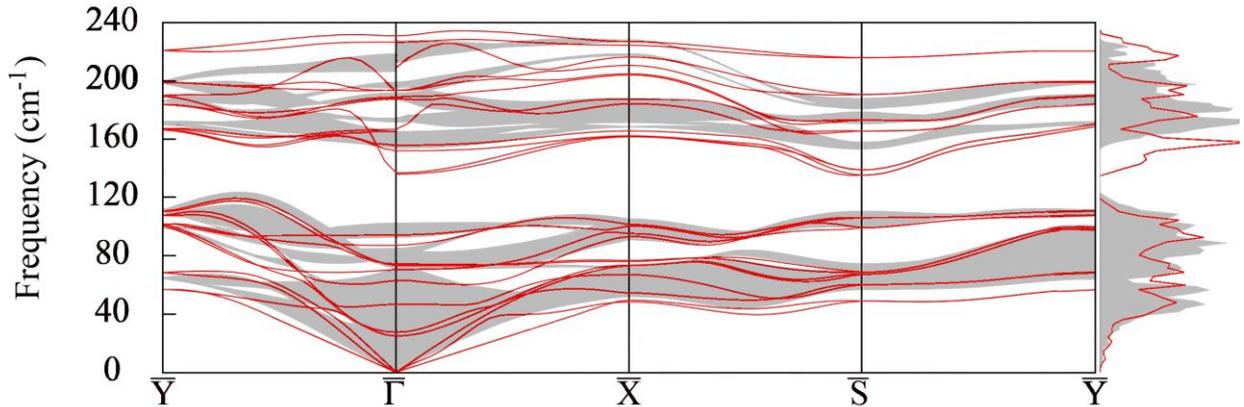

**Figure 1.** Calculated phonon spectra and phonon density of states (inset to the right side) of GeSe.

The topmost mode does not mix with bulk states and can be observed as true surface vibrations throughout the two-dimensional BZ. The lowest surface mode which starts at point **Γ**, as a shear-horizontal mode, is the Rayleigh wave (RW).
In addition, one can identify a group of surface states in the forbidden zone between two zones that are fully localized at the surface. All other surface states fall in the area of the bulk vibrations and can be classed as mixed states, when both the bulk and the surface atoms participate in vibrations.
Fig. 1 also shows the local density of states (LDOS), along with the bulk density. One can identify two regions in the LDOS for the surface layer. The main contribution into the first peak in the first area comes from the low-frequency modes located below the bulk states. The first and the last peaks in the second region are determined by the vibrations in the forbidden gap, and these have the localized character of vibrations. All other peaks are mixed vibrations. This indicates strong hybridization of the vibrational states of the surface and of the bulk.

**Conclusion**

In this paper we present the results of ab initio calculations of the phonon dynamics of the GeSe (001) surfaces. We demonstrate that the localized modes appear in the gaps of the bulk continuum as well as within the pockets of the bulk subbands for certain directions of symmetry. The projected phonon dispersion curves revealed the presence of a Rayleigh mode. A change in the surface force constants signifcantly modifies the qualitative features of the surface phonon spectra.

**References**

[1] Fritsch J., Schroder U., Physics Reports, 1999, **309**, 209.
[2] Panella V., Glebov A. L., Toennies J. P., Sebenne C., Eckl C., Adler C., Pavone P., Schroder U., Phys. Rev. B, 1999, **59** , 15772.
[3] Hsueh H. C., Vass H., Clark S. J., Ackland G. J., Crain J., Phys.Rev. B, 1995, **51**, 16750.
[4] Gonze X., Beuken J. M., Caracas R., Detraux F., Fuchs M., Rignanese M., Sindic L., Verstraete M., Zerah G., Jallet F., Comput. Mater. Sci., 2002, **25**, 478.
[5] Goedecker S., Teter M., Huetter J., Phys. Rev. B, 1996, **54**, 1703.
[6] Hartwigsen C., Goedecker S., Hutter J., Phys. Rev. B, 1998, **58**, 3641.
[7] Monkhorst H., Pack J., Phys. Rev. B,1976, **13**, 5188.